\documentclass[
aps,
prab,
reprint,
superscriptaddress,
amsmath,
amssymb,
floatfix]{revtex4-2}
\usepackage[utf8]{inputenc}
\usepackage[T1]{fontenc}
\usepackage{graphicx}
\usepackage{bm}

\DeclareUnicodeCharacter{00B0}{\ensuremath{^\circ}}      
\DeclareUnicodeCharacter{2212}{\ensuremath{-}}           
\DeclareUnicodeCharacter{207B}{\textsuperscript{-}}      
\DeclareUnicodeCharacter{00B9}{\textsuperscript{1}}      
\DeclareUnicodeCharacter{00B2}{\textsuperscript{2}}      
\DeclareUnicodeCharacter{00B3}{\textsuperscript{3}}      
\DeclareUnicodeCharacter{2070}{\textsuperscript{0}}      
\DeclareUnicodeCharacter{2081}{\textsubscript{1}}        
\DeclareUnicodeCharacter{2082}{\textsubscript{2}}        
\DeclareUnicodeCharacter{03BC}{\ensuremath{\mu}}         
\DeclareUnicodeCharacter{03A9}{\ensuremath{\Omega}}      
\DeclareUnicodeCharacter{03B5}{\ensuremath{\varepsilon}} 
\DeclareUnicodeCharacter{0393}{\ensuremath{\Gamma}}      
\DeclareUnicodeCharacter{0394}{\ensuremath{\Delta}}      
\DeclareUnicodeCharacter{03B4}{\ensuremath{\delta}}      
\DeclareUnicodeCharacter{2248}{\ensuremath{\approx}}     
\DeclareUnicodeCharacter{00B1}{\ensuremath{\pm}}         
\DeclareUnicodeCharacter{2273}{\ensuremath{\gtrsim}}     
\DeclareUnicodeCharacter{2272}{\ensuremath{\lesssim}}    
\DeclareUnicodeCharacter{221D}{\ensuremath{\propto}}     

\begin{document}

\title{Design Study of an Endless RF Phase Shifter Using Ferroelectric Capacitors}

\author{I. Ben-Zvi}
\email{Ilan.Ben-Zvi@StonyBrook.edu}
\affiliation{Physics and Astronomy Department, Stony Brook University, NY USA}
\date{\today}

\begin{abstract}
An endless radio-frequency (RF) phase shifter is designed using
ferroelectric capacitors. An endless phase winds continuously and
without bound while every control voltage executes a bounded periodic
cycle. A new scheme for an endless RF phase shifter is proposed and
studied in the framework of an equivalent circuit. Besides the endless
property, this phase shifter offers an exceptionally low insertion-loss,
high-speed and high-power capability. The device may perform as an exact
frequency translator whose offset is locked to the bias repetition
frequency. Optimized realizations are given at 400 and 800 MHz with full
conductor-loss accounting: 0.19 and 0.28 dB average insertion-loss. A
phase advance rate of a few microseconds per turn is limited only by the
bias electronics, and power capability on the order of 100 kW may be
achieved through stacked-wafer construction. An instantaneous bandwidth
of approximately 2.5\% is loss-limited rather than mechanism-limited.
The design surpasses the best endless phase shifter in the literature,
the rotary-field ferrite device, in speed, loss, and power, and improves
on the serrodyne ferroelectric frequency translator in efficiency and
drive electronics requirements.
\end{abstract}

\maketitle

\section{Introduction}
There is a recurring need in RF and microwave systems for a phase
actuator with unbounded range. This communication addresses this need by
offering an endless (continuous unbound phase range) RF phase shifter
based on ferroelectric capacitors in resonant structures. It will be
demonstrated that this device has a low insertion-loss, high linearity,
average-power capability and fast phase-advance speed. To anchor the
discussion to a particular application, the example is a magnetron
locked in phase to a superconducting accelerator cavity. A magnetron
wanders in frequency, and the phase lock must track the integrated
frequency error --- a phase that may grow above 360° in either
direction. A conventional phase shifter, however wide its range, must
eventually reset; the serrodyne endless phase shifter of \cite{benzvi2025hpffm},
developed for the same application, resets through \(2\pi\) requiring a
very fast flyback and some loss in efficiency. A device that never
resets --- an endless phase shifter --- removes this problem at its
root.

Because the useful phase of any such device is defined modulo 360°, the
phase-range-per-loss figure of merit customary for ordinary phase
shifters carry no information here: the relevant measure of RF
performance is the insertion-loss itself. Two measures are applied, the
phase-averaged value over one turn, \({IL}_{avg}\) and the worst
instantaneous value \({IL}_{\max}\).

The tunable element is the bulk ferroelectric capacitor developed for
accelerator applications \cite{kazakov2010fast,kanareykin2005lowloss,kanareykin2006fast,nenasheva2010lowloss}: a BST(M) ceramic wafer assuming a
relative permittivity swings from \(\varepsilon_{2}\  = \ 130\) at zero
bias to \(\varepsilon_{1}\  = \ 96.4\) at a field of 8 MV/m, with
intrinsic response time of order $10^{-10}$ s and a loss tangent
\(\delta\  = \ 1.0\  \times \ 10^{- 3}\) at 400 MHz. The same wafers,
stacked for voltage division and heat removal, are the basis of 100
kW-class fast reactive tuners \cite{benzvi2024frt,benzvi2025tuners}, so the device inherits a
demonstrated path to high power. These three properties --- speed, low
loss, and power --- motivate a ferroelectric realization of the endless
function. The design study of this endless ferroelectric phase shifter
is carried out using an equivalent circuit model. An electromagnetic
simulation in three dimensions is beyond the scope of this work, but
necessary for a complete validation of the concept.

\section{Principle of operation}
Endless operation is a consequence of a trajectory of the reflection
coefficient in the complex plane (or the Smith chart). If the trajectory
encloses the vortex, the point where the reflection coefficient
vanishes, (\(\Gamma\  = \ 0\)), and returns to its departure point, the
phase changed by ±360°. A distortion of the trajectory that maintains
this enclosure affects the insertion-loss but not the phase of the full
circuit. A low loss trajectory is one that stays close to the
\(|\Gamma| = 1\) boundary.

A periodic drive returns the controls exactly to their starting values
each period; the reflection coefficient, being a single-valued function
of the controls, then returns exactly to its starting value, and the
accumulated phase per branch control-period is forced to be 360° times
an integer --- the winding number --- with no calibration entering the
statement.

At any point along the trajectory the control may change direction or
shift the phase as a function of time according to what its application
requires, however such a phase change may exceed 360°. Thus, the endless
phase shifter operates like any voltage-controlled phase shifter except
that the phase range is unlimited.

A single resonator can't cover a range of 360°. Similarly, a circuit of
two resonators sharing the same control voltage of their resonant
frequency would not do it.

Winding the phase of the reflection coefficient at least once requires
two frequency-controlled resonators driven differently. In this work,
series connection of two parallel resonators (\textquotesingle tanks')
is adopted, defined as a ``branch''. One resonator is designated as
``gate'' and the other as ``phase''. The forward transit of one must
occur at full port coupling and the return transit at (nearly) zero
coupling. The controls must therefore modulate alternately the coupling
of the branch as well as its phase tuning. In the plane of the two
controls, the closed trajectory must enclose a singular point of the
branch's reflection coefficient. Such a point is precisely defined: it
is a control setting
\(\left( {\varepsilon_{p}}^{{^\circ}},\ {\varepsilon_{g}}^{{^\circ}} \right)\)
at which the branch impedance \(Z_{br}\) equals the port impedance
\(Z_{0}\) exactly, \(Z_{br}\  = \ Z_{0}\), so that the incident wave is
perfectly absorbed and \(\Gamma\  = \ 0\). At a match point the phase of
Γ is undefined and winds by ±360° on any small loop around it --- a
phase vortex. Since the reflection coefficient is a single-valued
function of the controls, the phase accumulated over any closed control
cycle equals 360° times the net number of vortices it encloses, counted
with sign: a topological invariant, independent of waveform shapes,
amplitudes, and component drift. Matching imposes two real conditions on
the two controls, so vortices are isolated points. For the 400 MHz
design of Section 5 the enclosed vortex lies at
\(\left( {\varepsilon_{p}}^{{^\circ}},\ {\varepsilon_{g}}^{{^\circ}} \right)\)
= (98.6, 125.9), marked in Fig. 2. An opposite-sign partner vortex lies
outside the operating rectangle at (93.9, 134.4). The operating
trajectory itself never approaches the vortex --- the deepest coupling
anywhere on the cycle keeps \(|\Gamma|\  \geq \ 0.977\) --- the cycle
merely encircles it. The order of operations is thus the essence of the
device: traversing the same four control moves in a sequence that does
not enclose the vortex (for instance, rewinding the phase capacitor at
the same gate state in which it swept) yields exactly zero; traversing
the enclosing sequence in reverse reverses the sign of the winding.

\begin{figure}[htbp]
\centering
\includegraphics[width=1.00\columnwidth]{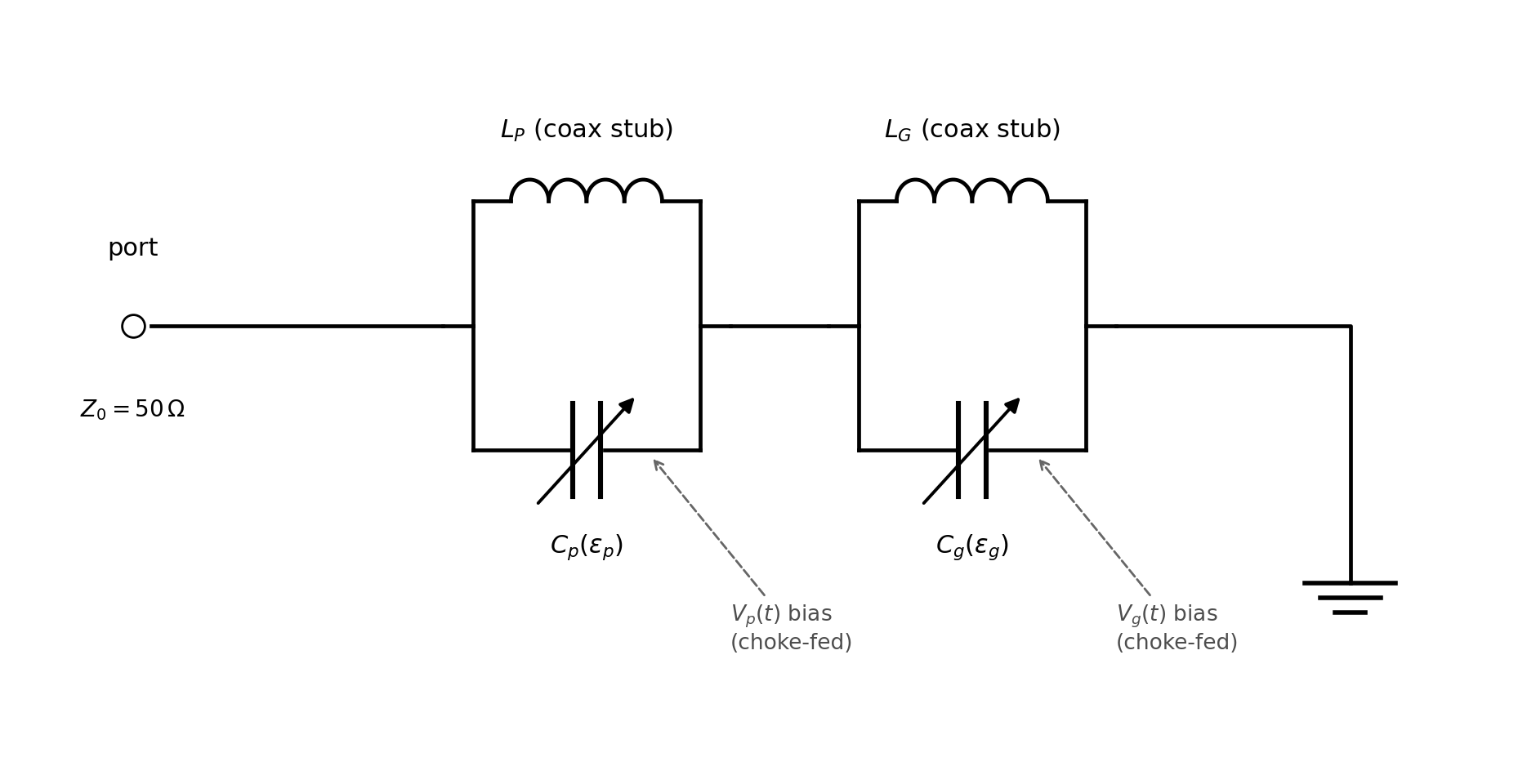}
  \caption{One branch of the endless phase shifter. Bias reaches the floating wafers through RF chokes.}
  \label{fig:1}
\end{figure}

Figure 1 shows the adopted branch and Fig. 2 its control cycle. Each
resonator blocks the branch at one end of its own permittivity range:
the gate tank at its resonance
\(\varepsilon_{g,res}\  = \ \varepsilon_{2}\), where its parallel
impedance is large and real and the branch draws essentially no current,
and the phase tank at \(\varepsilon_{pa}\  = \ \varepsilon_{1}\), placed
by design at the low end of the sweep. The cycle passes the blocking
baton between the two: on edge A (unblock) the gate opens while the
phase tank still self-blocks the branch, so the move is silent (−0.8°);
on edge B the phase tank releases the branch into and through the
full-coupling resonance (−192.1°); on edge C the gate reclaims it
(−167.8°); on edge D the phase capacitor rewinds behind the open gate
(−0.1°). All the phase is generated on the two edges where the branch
passes between blocked and coupled through different blockers, and the
four contributions sum to −360.00° exactly.

The permittivities are controlled by bias voltages applied to the
ferroelectric capacitors. The direction of the control cycle can be
reversed to unwind the phase, stopped to maintain a constant phase or
modulated about any phase.

A single branch already performs as an endless phase shifter, and that
may already be a useful device for various applications, such as the
distribution of high-level RF power from a single klystron to multiple
RF accelerator cavities at a fixed arbitrary phase. However, the cycle
of the branch contains two sections in which the phase advance is very
small, and that is not desirable in a phase shifter where arbitrary fast
modulation may be present at any time. This limitation is removed when
the endless phase shifter comprises two branches. The second branch
executes with a quarter cycle lag; thus, one branch is always winding
while the other is silent. Each branch completes one full cycle per 360°
of phase that the branch contributes, and the device output advances
720° per bias period with its phase being a smooth function of the
control. Driven periodically at a per-branch repetition frequency
\(f_{bias}\), this winding rate produces a frequency offset
\(\Delta f\  = \ 2f_{bias}\).

\begin{figure}[htbp]
  \centering
  \includegraphics[width=\columnwidth]{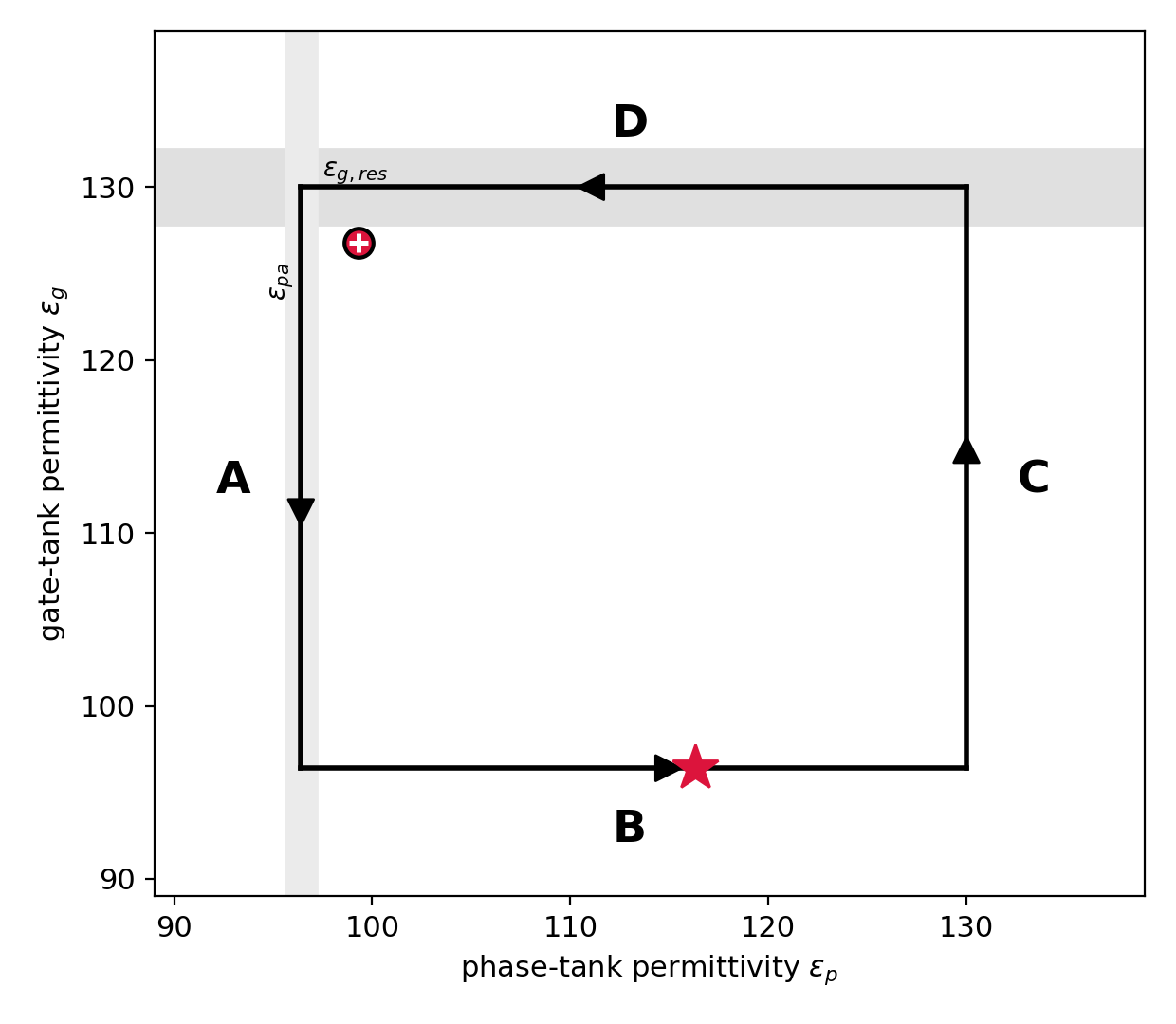}
  \caption{The control cycle of a single branch in the
\(\left( \varepsilon_{p},\ \varepsilon_{g} \right)\) plane.}
  \label{fig:2}
\end{figure}

In Figure 2, at Segment A the gate leaves its resonance and opens
while the phase tank, parked beside its own resonance
\(\varepsilon_{pa}\) (light vertical band), still blocks the
branch. At Segment B the phase
tank releases the branch into and through the deepest coupling (star). At Segment C the gate returns to its
resonance \(\varepsilon_{g,res}\) (gray horizontal band) and
opens the branch. Finally, at Segment D the phase
capacitor rewinds behind the open branch. The star marks the
deepest coupling reached on the sweep, at -180 degrees. The filled circle marks the
phase vortex, the perfect-match point \(\Gamma\  = \ 0\)
defined in the text, which the operating trajectory never
approaches. The cycle encloses this vortex once, so the phase per branch
cycle is −360° exactly (winding number \(w\  = \ 1\)); reversing
the sequence reverses the sign.

\section{Circuit model}
All impedances are evaluated at the fixed operating angular frequency
\(\omega\  = \ 2\pi f\) and normalized results are quoted for a port
impedance \(Z_{0}\  = \ 50\ \Omega\). The ferroelectric material
properties \cite{kazakov2010fast,kanareykin2005lowloss,kanareykin2006fast,nenasheva2010lowloss,fox1947} are taken as: relative permittivity ε between
\(\varepsilon_{2}\  = \ 130\) (zero bias) and
\(\varepsilon_{1}\  = \ 96.4\) (8 MV/m), with center value
\(\varepsilon_{c}\  = \ \sqrt{\varepsilon_{1}\varepsilon_{2}}\  = \ 112.0\),
and the loss tangent δ scales with frequency as

\begin{equation}
  \delta(f)\  = \ \delta_{400}\ \left( \frac{f}{400\ MHz} \right)^{0.63}\ ,\ \ \ \ \ \ \delta_{400}\  = \ 1.0\  \times \ 10^{- 3}\
  \tag{1}\label{eq:1}
\end{equation}

The quoted values \(\varepsilon_{1}\  = \ 96.4\),
\(\varepsilon_{2}\  = \ 130\), and
\(\delta_{400}\  = \ 1.0\  \times \ 10^{- 3}\) are representative of the
BST(M) compositions of \cite{kazakov2010fast,kanareykin2005lowloss,kanareykin2006fast,nenasheva2010lowloss} and may vary from sample to sample;
the design procedure takes the measured pair
\(\left( \varepsilon_{1},\ \varepsilon_{2} \right)\) and \(\delta\) as
inputs, and material properties are referred to symbolically hereafter.

The capacitor quality factor is \(Q\  = \ 1/\delta\  = \ 1000\) at 400
MHz. Each ferroelectric capacitor is a parallel-plate wafer of area
\(A_{x}\) and thickness \(h_{x}\) (\(x\  = \ p\) for the phase element,
\(x\  = \ g\) for the gate), with \(\varepsilon_{0}\) the vacuum
permittivity:

\begin{equation}
  C_{x}(\varepsilon)\  = \ \frac{\varepsilon\ \varepsilon_{0}\ A_{x}}{h_{x}}\ ,\ \ \ \ \ \ x\  \in \ \{ p,\ g\}
  \tag{2}\label{eq:2}
\end{equation}

The reactance is
\(X_{Cx}(\varepsilon)\  = \ 1/\omega C_{x}(\varepsilon)\ \) , and its
impedance:

\begin{equation}
  Z_{Cx}(\varepsilon)\  = \ \frac{\delta\  - \ j}{\omega\ C_{x}(\varepsilon)}\  = \ X_{Cx}(\varepsilon)\ (\delta\  - \ j)
  \tag{3}\label{eq:3}
\end{equation}

The two inductors of each branch are shorted coaxial copper stubs of
inner and outer radii \(a\  = \ 2\ cm\) and \(b\  = \ 4.25\ cm\) and
lengths \(l_{P},\ l_{G}\), evaluated with the exact lossy-line input
impedance, including the resistance of both annular end faces:

\begin{equation}
  Z_{Lx}\  = \ R_{end}\  + \ Z_{c}\ \frac{R_{end}\  + \ Z_{c}\ tanh\left( \gamma l_{x} \right)}{Z_{c}\  + \ R_{end}\ tanh\left( \gamma l_{x} \right)}
  \tag{4}\label{eq:4}
\end{equation}

\begin{widetext}
\begin{equation}
  Z_{c}\  = \ \sqrt{\frac{R'\  + \ j\omega L'}{j\omega C'}}\ ,\ \ \ \gamma\  = \ \sqrt{(R'\  + \ j\omega L')\ j\omega C'}\ ,\ \ \ R'\  = \ \frac{R_{s}}{2\pi}\left( \frac{1}{a}\  + \ \frac{1}{b} \right)\ ,\ \ \ R_{end}\  = \ \frac{R_{s}}{2\pi}\ ln\frac{b}{a}\ ,\ \ \ R_{s}\  = \ \sqrt{\pi\ f\ \mu_{0}\ \rho}
  \tag{5}\label{eq:5}
\end{equation}
\end{widetext}

Here \(\rho\  = \ 1.72\  \times \ 10^{- 8}\ \Omega \cdot m\) is the
copper resistivity, \(\mu_{0}\) the vacuum permeability, \(R_{s}\) the
surface resistance, \(R'\) the body resistance per unit length,
\(R_{end}\) the end-cap resistance, and \(L'\), \(C'\) the inductance
and capacitance per unit length of the coax; the expression reduces to
the lumped-inductor model at small electrical length. Each parallel
resonator then has impedance

\begin{equation}
  Z_{X}(\varepsilon)\  = \ \left\lbrack \frac{1}{Z_{LX}}\  + \ \frac{1}{Z_{CX}(\varepsilon)} \right\rbrack^{- 1}\ ,\ \ \ \ \ \ X\  \in \ \{ P,\ G\}
  \tag{6}\label{eq:6}
\end{equation}

and the branch impedance as a function of the two permittivities --- the
central circuit equation of the device --- is simply the series chain of
the two tanks:

\begin{equation}
  Z_{br}\left( \varepsilon_{p},\ \varepsilon_{g} \right)\  = \ Z_{P}\left( \varepsilon_{p} \right)\  + \ Z_{G}\left( \varepsilon_{g} \right)
  \tag{7}\label{eq:7}
\end{equation}

The central values of the branch's two capacitors are set by optimizing
the insertion-loss.

The complete endless phase shifter comprises two such branches in
parallel; with \(Y\) the total admittance, the reflection coefficient
and insertion-loss are

\begin{widetext}
\begin{equation}
  Y\  = \ \frac{1}{Z_{br,1}}\  + \ \frac{1}{Z_{br,2}}\ ,\ \ \ \ \ \ \Gamma\  = \ \frac{1\  - \ Z_{0}Y}{1\  + \ Z_{0}Y}\ ,\ \ \ \ \ \ IL\  = \  - 20\ \log_{10}\ |\Gamma|
  \tag{8}\label{eq:8}
\end{equation}
\end{widetext}

The element values \(C_{g}(\varepsilon_{c})\) and
\(C_{p}(\varepsilon_{c})\) are determined through a constrained
optimization: minimize the phase-weighted \({IL}_{avg}\) over the cycle,
subject to the topological constraint that the swept rectangle [$\varepsilon_{1}$,
$\varepsilon_{2}$]² encloses the perfect-match vortex.

At any point in the optimization the values of the inductances are fixed
by the gate-tank resonance (branch open) and the phase-tank resonance
(branch self-blocked)

\begin{equation}
  \omega^{2}L_{G}C_{g}\left( \varepsilon_{g,res} \right)\  = \ 1\ \  \Rightarrow \ \ \varepsilon_{g,res}\  = \ \varepsilon_{2}
  \tag{9}\label{eq:9}
\end{equation}

\begin{equation}
  \ \ \ \omega^{2}L_{P}C_{p}\left( \varepsilon_{pa} \right)\  = \ 1\ \  \Rightarrow \ \ \varepsilon_{pa}\  = \ \varepsilon_{1}
  \tag{10}\label{eq:10}
\end{equation}

The full-coupling position \(\ \ Im\ Z_{br}\  = \ 0\ \ \)emerges from
the optimization procedure. Its significance is being the most
measurable feature the device has. On a network analyzer it is simply
the $|S_{11}|$ dip during a slow phase-capacitor sweep with
the gate open, and the bias voltage at which it occurs anchors the
bias-to-ε calibration of the ferroelectric material.

\section{Anatomy of one control
cycle}
\begin{figure*}[htbp]
  \centering
  \includegraphics[width=0.95\textwidth]{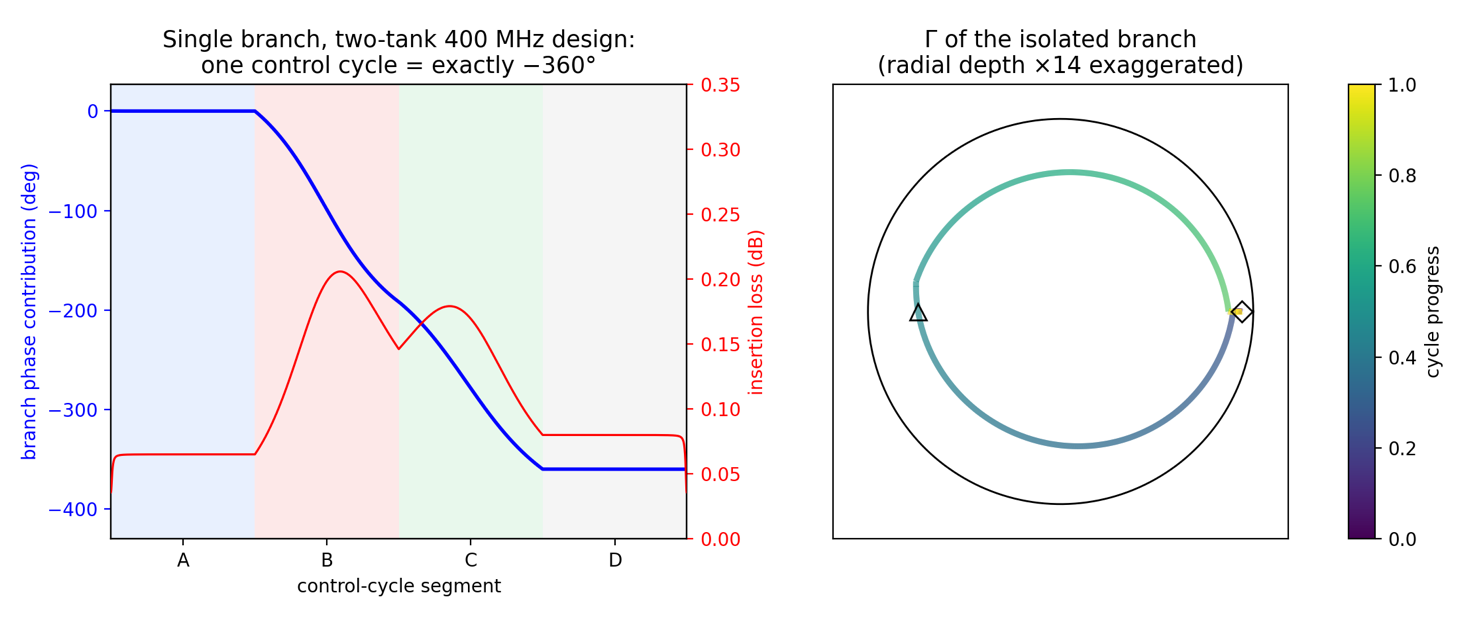}
  \caption{One isolated branch of the 400 MHz design over its
four-segment control cycle (conductor losses included; segment durations
drawn equal for clarity). Segments A (unblock), B (sweep), C (block),
and D (rewind, silent). Left: the phase contributions and the
insertion-loss of a single branch vs the control cycle. Right: the
reflection coefficient winds once around the chart (radial depth
exaggerated by a factor of 14). the triangle marks the full-coupling
point Im Z\_br = 0, where the reflection phase crosses −180°; the
diamond marks the park point, at which the cycle begins and ends and
where the reflection coefficient dwells during the rewind while the
phase capacitor recharges behind the open gate.}
  \label{fig:3}
\end{figure*}

Figure 3 presents a single branch cycle. Two features deserve emphasis.
First, the near-equal division between the sweep and block contributions
(−192.1° and −167.8°) is not imposed but emerges from the loss
optimization: sharing the turn between the two working moves at matched
depth produces the flattest, lowest-loss trajectory. Second, the extreme
asymmetry between the two gate transits --- the block generates half the
turn while the unblock generates almost nothing --- is the winding
mechanism made visible. The transits occur against different phase-tank
backdrops: the unblock at \(\varepsilon_{p}\  = \ \varepsilon_{1}\),
where the phase tank sits at its own resonance and holds the branch open
by itself, and the block at \(\varepsilon_{p}\  = \ \varepsilon_{2}\),
where the phase tank does not block. This demonstrates numerically the
argument of Section 2 which describes the principle of the endless phase
shifter.

\section{Optimized realizations at 400 and 800
MHz}
The circuit parameters --- the two capacitor impedance levels and the
phase-tank resonance placement \(\varepsilon_{pa}\) --- were optimized
at each frequency separately. Table 1 gives the result. Note that the
phase shifter described is a one-port reflective device, and an
application requiring separation of the incident and reflected waves
requires an additional RF component, such as magic Tee or circulator.
Each branch contains exactly two elements of copper, both short high-Q
stubs, and the conductor-loss penalty is only 20\% of the total at 400
MHz and 14\% at 800 MHz. The parked branch has a residual susceptance
below $10^{-3}$ of the port admittance. While this holds in the idealized
equivalent circuit model, with a 3D electromagnetic model a residual
admittance may materialize, anticipating a fixed trim stub to cancel it
in a future EM study. The gate resonates at
\(\varepsilon_{g}\  = \ \varepsilon_{2}\), i.e., at zero bias, so a
drift margin implied by this choice is quantified below.

The optimum capacitance levels are decided by the conductors: large
capacitances (small tank reactances) shorten the stubs and shrink their
resistance, favoring the 10--40 pF values of Table 1, while the retreat
toward still larger values is halted by the end-cap-dominated
degradation of stub quality factor at very small reactance.

\begin{table*}[htbp]
  \caption{Final circuit realizations of the endless phase shifter
at a 50 Ω port. Ferroelectric material and coax resistance models per
\cite{kazakov2010fast,kanareykin2005lowloss,kanareykin2006fast,nenasheva2010lowloss,benzvi2024frt,benzvi2025tuners,fox1947}; for high-power operation each wafer is replaced by
\(N_{w}\) stacked wafers \cite{benzvi2024frt,benzvi2025tuners}. Dissipations are fractions of
incident RF power under the constant-frequency drive of Section 6.}
  \label{tab:1}
  \begin{ruledtabular}
  \begin{tabular}{p{0.30\textwidth} p{0.30\textwidth} p{0.30\textwidth}}
  \textbf{Quantity} & \textbf{400 MHz} & \textbf{800 MHz} \\
  \hline
  \(C_{p}\) at
\(\varepsilon\  = \ \varepsilon_{1}\ /\ \varepsilon_{c}\ /\ \varepsilon_{2}\)
& 23.9 / 27.7 / 32.2 pF & 11.9 / 13.8 / 16.0 pF \\
  \(L_{P}\) stub (\(a\  = \ 2\ cm\), \(b\  = \ 4.25\ cm\)) & 16.7 Ω,
\(l\  = \ 4.21\ cm\), \(Q\  = \ 4024\) & 16.7 Ω, \(l\  = \ 2.12\ cm\),
\(Q\  = \ 4310\) \\
  \(C_{g}\) at
\(\varepsilon\  = \ \varepsilon_{1}\ /\ \varepsilon_{c}\ /\ \varepsilon_{2}\)
& 21.5 / 25.0 / 29.0 pF & 10.7 / 12.4 / 14.4 pF \\
  \(L_{G}\) stub & 13.7 Ω, \(l\  = \ 3.52\ cm\), \(Q\  = \ 3825\) & 13.8
Ω, \(l\  = \ 1.77\ cm\), \(Q\  = \ 3973\) \\
  Insertion-loss \({IL}_{avg}\) / \({IL}_{\max}\) & 0.19 / 0.23 dB & 0.28/
0.33 dB \\
  Segment shares (sweep / block) & −192.1° / −167.8° & −192.2° /
−167.7° \\
  Dissipation, time-avg per branch & \(C_{p}\) 0.92\% (2.9\%),
\(C_{g}\) 0.81\% (2.4\%) & see Section 6 \\
  Peak RF field amplitude at 100 kW (for \(N_{w}\  = \ 4\); ∝
\(\sqrt{P_{inc}}/N_{w}\)) & \(C_{p}\) 4.6, \(C_{g}\) 4.3 MV/m &
\(C_{p}\) 4.6, \(C_{g}\) 4.3 MV/m \\
  \end{tabular}
  \end{ruledtabular}
\end{table*}

The placement of the gate resonance at the range end maximizes the
gate\textquotesingle s blocking action but leaves one-sided margin
against drift of the resonance. For example, at 400 MHz, assume
retreating the resonance to
\(\varepsilon_{g,res}\  = \ \varepsilon_{2}\  - \ m\), recomputing
\(L_{G}\) from Eq. (9) for each placement and holding all other
parameters fixed. That buys a two-sided drift margin \(m\) at a
reasonable loss penalty, just 7\% in \({IL}_{avg}\) at \(m\  = \ 3\),
and 17\% at \(m\  = \ 6\).

The instantaneous bandwidth of the device, with applied 100 kHz
modulation, is calculated here by two criteria: (a) survival of the
phase per full device period remaining exactly −720°, and (b) the
phase-averaged insertion-loss does not exceed twice its design value,
with the worst-case loss monitored alongside.

\begin{table*}[htbp]
  \caption{Instantaneous bandwidth of the adopted designs.}
  \label{tab:2}
  \begin{ruledtabular}
  \begin{tabular}{p{0.30\textwidth} p{0.30\textwidth} p{0.30\textwidth}}
  \textbf{Criterion} & \textbf{400 MHz} & \textbf{800 MHz} \\
  \hline
  Winding preserved (topological) & beyond −2\% to +1.9\% & beyond −2\% to
+2.2\% \\
  \({IL}_{avg}\  \leq \ 2 \times\) design value & −1.37\% to +1.18\% (10.2
MHz) & −1.51\% to +1.35\% (22.9 MHz) \\
  \end{tabular}
  \end{ruledtabular}
\end{table*}

\section{Power dissipation under constant-frequency-shift operation}
It is useful to examine the performance of the endless phase shifter in
view of a practical application. In applying phase locking of a
magnetron to a master clock, the phase shifter follows a servo command
to correct the phase. The winding rate is the derivative of the phase,
which is the instantaneous frequency error of the magnetron. The
magnetron's frequency varies over a large range with fast fluctuation. A
constant frequency shift is analyzed here as the worst-case sustained
slew. It is also the operating mode of the device as a frequency
translator. In a phase control scenario, the phase error would advance
(or retard) the clock generating the control voltages. A constant rate
frequency translation \(\Delta f\) requires the total phase to be
exactly linear in time, which fixes the waveforms completely. A
reclocking transformation is necessary to generate a look-up table (LUT)
that includes the device phase vs the four permittivity values of the
ferroelectric capacitors and permittivity vs bias voltage. The unblock
and rewind of each branch then execute while the partner ramps, and
their vanishing contributions are absorbed as momentary adjustments of
the partner\textquotesingle s ramp rate. All the control voltages are
slaved to the master clock.

\begin{figure*}[htbp]
  \centering
  \includegraphics[width=0.72\textwidth]{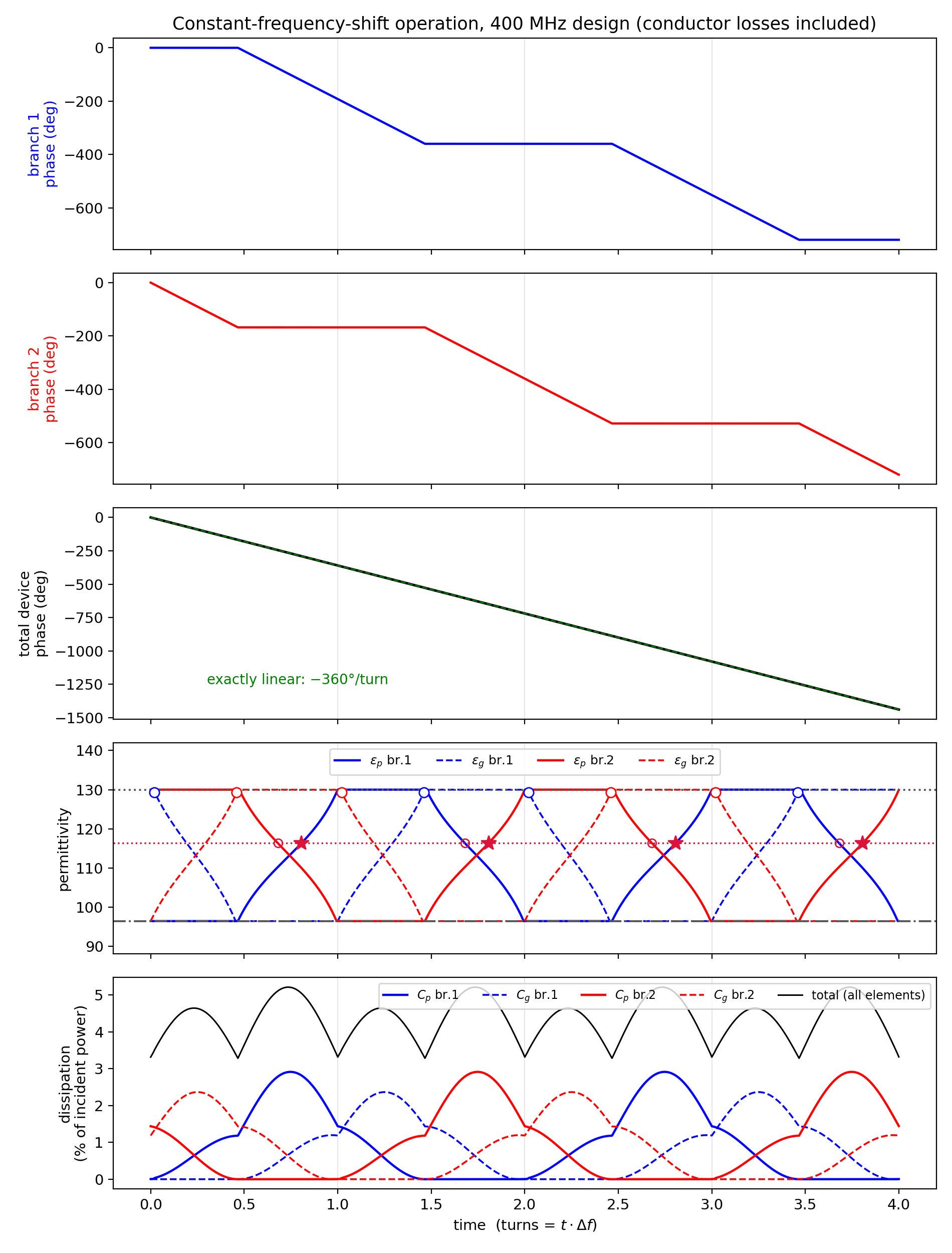}
  \caption{Constant-frequency-shift operation of the 400 MHz design
(conductor losses included), synthesized by phase reclocking, with the
unblock and rewind spread over their full windows. Time is in turns, 10
μs per turn at a frequency translation of 100 kHz. Panels, top to
bottom: the two branch phases; the total device phase against the
−360°/turn line; the four permittivity waveforms; the dissipation in the
capacitors as percentage of the incident power.}
  \label{fig:4}
\end{figure*}

Figure 4 shows the result: the per-branch
staircases, the linear total, the four permittivity waveforms with the
two resonant permittivities marked, and the dissipation in every
ferroelectric capacitor as a fraction of the incident RF power.
Additional information is Figure 4 includes horizontal lines marking the
gate-tank resonance \(\varepsilon_{g,res}\  = \ \varepsilon_{2}\), the
phase-tank resonance \(\varepsilon_{pa}\  = \ 96.4\) and the
deepest-coupling permittivity \({\varepsilon_{p}}^{*}\  = \ 116\).
Larger circles mark the instants at which each gate permittivity reaches
or leaves its blocking resonance, stars the working transits of
\(\varepsilon_{p}\) through the deepest coupling, and smaller circles
the silent recrossings during rewind behind the open gate.

The dissipated power of each capacitor is the loss tangent times its
reactive power, which in design quantities reads

\begin{widetext}
\begin{equation}
  P_{Cx}\  = \ \frac{1}{2}\ \delta(f)\ \omega\ C_{x}\left( \varepsilon_{x} \right)\ {V_{x}}^{2}\  = \ \frac{1}{2}\ \delta(f)\ \omega\ \varepsilon_{x}\varepsilon_{0}\ A_{x}h_{x}\ {E_{x}}^{2}\ ,\ \ \ \ \ \ x\  \in \ \{ p,\ g\}
  \tag{11}\label{eq:11}
\end{equation}
\end{widetext}

where \(V_{x}\) is the RF voltage amplitude across capacitor \(x\) and
\(E_{x}\  = \ V_{x}/h_{x}\) the RF field amplitude in its wafer. The
field is related to electric breakdown or generation of non-linearities
\cite{benzvi2024frt}. Its peak values over the cycle are therefore given in Table 1,
scaling as the square root of the incident power and inversely with the
number of wafers \(N_{w}\). The exact time profiles of Fig. 4 are
evaluated from the full circuit model, and the sum of all element
dissipations adds to the total absorbed fraction
\(1\  - \ {|\Gamma|}^{2}\) .

The field itself is evaluated directly from the design and the incident
power: with the parked branch open, the port voltage is
\(\sqrt{2\ Z_{0}\ P_{inc}}\ |1\  + \ \Gamma|\) with
\(1\  + \ \Gamma\  = \ 2Z_{br}/\left( Z_{br}\  + \ Z_{0} \right)\), and
the branch current times the tank impedance is the capacitor voltage,
giving

\begin{widetext}
\begin{equation}
  E_{x}\  = \ \frac{2}{h_{x}}\ \sqrt{2\ Z_{0}\ P_{inc}}\ \frac{|Z_{X}\left( \varepsilon_{x} \right)|}{|Z_{br}\left( \varepsilon_{p},\ \varepsilon_{g} \right)\  + \ Z_{0}|}\ ,\ \ \ \ \ \ x\  \in \ \{ p,\ g\}
  \tag{12}\label{eq:12}
\end{equation}
\end{widetext}

with \(Z_{X}\) and \(Z_{br}\) evaluated from Eqs. (6)--(7) at the
instantaneous permittivities, allowing the field evaluation along the
cycle. Substituted into Eq. (11) it returns the dissipated power of each
capacitor; for a stack the per-wafer field is reduced by factor
\(N_{w}\).

Under the constant-rate drive the time average equals the phase average
identically, so the heat load is set by the average insertion-loss
\({IL}_{avg}\) of Table 1. The periodic insertion-loss variation over
the turn produces AM sidebands at harmonics of \(\Delta f\), which would
be flattened by the time response of a superconducting cavity.

Each capacitor may be realized as a stack of \(N_{w}\) wafers as in the
construction of \cite{benzvi2024frt,benzvi2025tuners}: the wafers are in series for the RF but
biased in parallel through the interleaved copper spacers, which also
carry the cooling. Holding the RF capacitance and the wafer thickness
\(h\  = \ 0.5\ mm\) fixed requires each wafer\textquotesingle s area to
grow as \(N_{w}A_{1}\), so the dissipation per wafer falls as
\(1/N_{w}\) while the cooled area grows as \(N_{w}\) --- the wafer
temperature rise falls as \(1/{N_{w}}^{2}\). With the thermal model of
\cite{benzvi2025tuners} (uniform heating, both faces cooled, average rise
\(\Delta T\  = \ P_{C}h\ /\ 12KA_{tot}\), thermal conductivity
\(K\  \approx \ 7\ W\ m^{- 1}K^{- 1}\)), the busiest capacitor at 100 kW
incident runs the temperature rises of Table 3. 

\begin{table*}[htbp]
  \caption{Temperature rise and drive power as function of the number of wafers at 400 MHz}
  \label{tab:3}
  \begin{ruledtabular}
  \begin{tabular}{p{0.14\textwidth} p{0.22\textwidth} p{0.24\textwidth} p{0.26\textwidth}}
  \(N_{w}\) \textbf{(wafers)} & \textbf{Wafer area (each)} & \textbf{ΔT (}\(C_{p}\)\textbf{), 100 kW} & \(P_{drive}\)\textbf{,} \(\Delta f\  = \ 100\ kHz\) \\
  \hline
  1 & 14.0 mm$^{2}$ & 391 K & 84 W \\
  2 & 28.0 mm$^{2}$ & 98 K & 337 W \\
  4 & 56.0 mm$^{2}$ & 24 K & 1.35 kW \\
  \end{tabular}
  \end{ruledtabular}
\end{table*}

Stacking divides this RF
voltage: the per-wafer field falls as \(1/N_{w}\). The price is paid by
the bias electronics: the drive sees the wafers in parallel, so the
capacitance presented to the supply grows as \({N_{w}}^{2}\), and for a
driver that dissipates the switched energy the drive power is bounded by

\begin{equation}
  P_{drive}\  \leq \ {N_{w}}^{2}\ \left( \langle C_{p}\rangle\  + \ \langle C_{g}\rangle \right)\ {V_{b}}^{2}\ \Delta f
  \tag{13}\label{eq:13}
\end{equation}

with \(V_{b}\  = \ 4\ kV\) the per-wafer bias swing and
\(\langle C_{p}\rangle,\ \langle C_{g}\rangle\) the mean capacitances
(each branch pair cycles once per two turns; the two branches together
give the rate \(\Delta f\)). At 100 kW incident,
\cite{kanareykin2005lowloss}\textquotesingle s engineering allowance of a 30 °C average rise
points to \(N_{w}\  = \ 4\) (or a modest further area increase) for the
phase capacitor, at a bias-driver bound of 1.35 kW for a sustained
\(\Delta f\  = \ 100\ kHz\).

\section{Comparison with endless phase shifters in the
literature}
Truly endless RF phase shifters --- devices whose phase winds
indefinitely under bounded controls form a short list.

Fox\textquotesingle s mechanical rotary-vane waveguide phase changer
(1947) winds the phase by twice the rotation angle of a half-wave plate
and is endless, low-loss, and power-capable, but mechanically slow
\cite{fox1947}. The modern best-in-class is the rotary-field ferrite phase
shifter of Boyd and Hord, in which a transverse quadrupole magnetic bias
field is rotated electrically in a ferrite-filled circular waveguide:
continuous, highly accurate 360° rotation, insertion-loss of roughly
0.5--1 dB across 2--20 GHz, switching of order 100 μs, and demonstrated
S-band units with 40 kW peak and 600 W average power \cite{hord1988,hord1989,boyd1995}. Vector
(I/Q) modulators are endless-capable and fast but, as passive networks,
carry ≳ 6 dB of intrinsic conversion loss and watt-class power limits,
with active versions trading loss for noise and linearity \cite{kebe2025}.
Finally, the endless phase shifter using ferroelectric capacitors and
serrodyne frequency shifting \cite{benzvi2025hpffm} shares many features with the
proposed device, including the high-power capability and speed but has a
lower efficiency. 

Against its closest true competitor, the serrodyne ferroelectric
\cite{benzvi2025hpffm}, the present design offers a better insertion-loss and avoids
the fast fly-back and its associated efficiency loss and lack of
complete phase coverage.

\section{Summary}
This work proposes a reflective RF phase shifter built from
ferroelectric capacitors in two gated branches of series-connected
parallel resonators. This phase shifter winds its phase endlessly, 360°
per branch per control cycle. The control sequence encloses the point
where the reflection coefficient vanishes (a vortex). Thus, the order of
the four control moves, not their shapes, is what generates the
function. Optimized realizations at 400 and 800 MHz, with the complete
coaxial conductor-loss model, deliver the full circle at 0.19 dB and
0.28 dB average insertion-loss per turn respectively, from branches of
only two resonators with no auxiliary element. The reclocked bias
waveforms produce a perfectly linear phase, so that a servo-driven lock
accumulates phase without slips, and a periodically driven device
translates frequency by exactly twice the per-branch bias repetition
rate. The ferroelectric material dissipation is distributed over four
capacitors, the highest averaging 0.92\% of the incident power. High
power operation is aided by the stacked-wafer tuner technology. The
instantaneous bandwidth is 2.5\% / 2.9\% at 400 / 800 MHz set by a
doubled average insertion-loss. It surpasses the best endless phase
shifter in the literature --- the rotary-field ferrite device
\cite{hord1989,boyd1995,kebe2025} --- by two orders of magnitude in speed and a factor of two
or more in loss at higher power; and it improves on the serrodyne
ferroelectric translator \cite{benzvi2025hpffm} by eliminating the flyback dead time
(better electrical efficiency), replacing the fast flyback drive with
control signals no faster than the working ramp, and lowering the
insertion-loss per 360° of phase advance.
\section*{AI Declaration}
During the preparation of this work, the author worked with Claude Fable
5 (Anthropic) during the month of July 2026 to survey literature,
develop the equivalent circuit of the phase shifter, calculate results
and produce graphics. After using this tool, the author reviewed and
edited the content as needed and takes full responsibility for the
content of this publication.

\bibliographystyle{apsrev4-2}
\bibliography{refs}

@misc{benzvi2025hpffm,
  author        = {I. Ben-Zvi and N. Shipman},
  title         = {High power fast frequency modulation},
  year          = {2025},
  eprint        = {2502.13312},
  archivePrefix = {arXiv},
}

@article{kazakov2010fast,
  author  = {S. Yu. Kazakov and S. V. Shchelkunov and V. P. Yakovlev and A. Kanareykin and E. Nenasheva and J. L. Hirshfield},
  title   = {Fast ferroelectric phase shifters for energy recovery linacs},
  journal = {Phys. Rev. ST Accel. Beams},
  volume  = {13},
  pages   = {113501},
  year    = {2010},
}

@inproceedings{kanareykin2005lowloss,
  author       = {A. Kanareykin and E. Nenasheva and S. Karmanenko and A. Dedyk and V. Yakovlev},
  title        = {Low-loss ferroelectric for accelerator applications},
  booktitle    = {Proc. PAC 2005},
  pages        = {4305--4307},
  organization = {IEEE},
  year         = {2005},
}

@article{kanareykin2006fast,
  author        = {A. Kanareykin and E. Nenasheva and V. Yakovlev and A. Dedyk and S. Karmanenko and A. Kozyrev and V. Osadchy and D. Kosmin and P. Schoessow and A. Semenov},
  title         = {Fast switching ferroelectric materials for accelerator applications},
  journal       = {AIP Conf. Proc.},
  volume        = {877},
  year          = {2006},
  eprint        = {physics/0612240},
  archivePrefix = {arXiv},
}

@article{nenasheva2010lowloss,
  author  = {E. Nenasheva and N. F. Kartenko and I. M. Gaidamaka and O. N. Trubitsyna and S. S. Redozubov and A. I. Dedyk and A. D. Kanareykin},
  title   = {Low loss microwave ferroelectric ceramics for high power tunable devices},
  journal = {J. Eur. Ceram. Soc.},
  volume  = {30},
  pages   = {395},
  year    = {2010},
}

@article{benzvi2024frt,
  author  = {I. Ben-Zvi and G. Burt and A. Castilla and A. Macpherson and N. Shipman},
  title   = {Conceptual design of a high reactive-power ferroelectric fast reactive tuner},
  journal = {Phys. Rev. Accel. Beams},
  volume  = {27},
  pages   = {052001},
  year    = {2024},
}

@article{benzvi2025tuners,
  author        = {I. Ben-Zvi and A. Macpherson and S. Smith},
  title         = {Detailed design and optimization of ferroelectric tuners},
  journal       = {Phys. Rev. Accel. Beams},
  volume        = {28},
  pages         = {093502},
  year          = {2025},
  eprint        = {2504.19825},
  archivePrefix = {arXiv},
}

@article{fox1947,
  author  = {A. G. Fox},
  title   = {An adjustable wave-guide phase changer},
  journal = {Proc. IRE},
  volume  = {35},
  pages   = {1489--1498},
  year    = {1947},
}

@article{hord1988,
  author  = {W. E. Hord},
  title   = {Design considerations for rotary-field ferrite phase shifters},
  journal = {Microwave Journal},
  volume  = {31},
  pages   = {105},
  year    = {1988},
}

@article{hord1989,
  author  = {W. E. Hord},
  title   = {Microwave and millimeter-wave ferrite phase shifters},
  journal = {Microwave Journal},
  volume  = {32},
  pages   = {81--94},
  year    = {1989},
}

@inproceedings{boyd1995,
  author    = {C. R. Boyd, Jr.},
  title     = {A latching ferrite rotary-field phase shifter},
  booktitle = {Proc. 1995 IEEE MTT-S Int. Microwave Symposium},
  address   = {Orlando, FL, USA},
  volume    = {1},
  pages     = {103--106},
  year      = {1995},
  doi       = {10.1109/MWSYM.1995.406087},
}

@article{kebe2025,
  author  = {M. Kebe and others},
  title   = {A survey of phase shifters for microwave phased array systems},
  journal = {Int. J. Circuit Theory Appl.},
  year    = {2025},
  doi     = {10.1002/cta.4298},
}

\end{document}